# The Science and Technology of Broad-Band X-ray Astrophysics

**from:** the Broad-Band X-ray Science Analysis Group (BBX-SAG)

**BBX-SAG Co-Chairs:** Chien-Ting Chen (MSFC), Kristin K. Madsen (GSFC) & Daniel Stern (JPL)
**Co-Authors:** Javier A. García (GSFC), Xiurui Zhao (Caltech), Alaa Salah Afifi (Kepler Space University), Adrianah Appleman (UAH), Steve Bongiorno (MSFC), Matteo Bachetti (Cagliari), Peter Boorman (MPE), Mckinley Brumback (Middlebury), Douglas Buisson (independent scientist), Filippo D'Ammando (Bologna), Kirtan Dixit (USRA/MSFC), Rahul Gupta (GSFC), Danielle Gurgew (USRA/MSFC), Gaurava K. Jaisawal (DTU Space, Technical University of Denmark), Phil Kaaret (MSFC), Renee Ludlam (Wayne State), Peter Maksym (MSFC), Labani Mallick (Mallick Space Research Center), Eric Miller (MIT), Kaya Mori (Columbia), Emanuele Nardini (INAF), Srikanth Panini Singam (USRA/MSFC), Mark Reynolds (OSU), M. Lynnie Saade (UAH), Aarran Shaw (Butler), Dustin Swarm (Univ. of Iowa), Nicholas Thomas (MSFC), Georgios Vasilopoulos (University of Athens), Dom Walton (Hertfordshire), Gaurav Waratkar (Caltech), Dan Wilkins (Ohio State University), Jooyun Woo (Clemson), Yi-Jung Yang (NYU Abu Dhabi)+++

## 1. Executive Summary

Much of the progress to our understanding of high-energy astrophysical phenomena made over the past decade has been driven by coordinated observations between multiple X-ray missions, extending the energy range of observations from the low, or soft energy range below ~10 keV of most focusing X-ray telescopes to the full breadth of the X-ray regime that includes the high, or hard energy range reaching an order of magnitude higher energies. Observations covering a wide range of energy enable the simultaneous study of thermal and non-thermal physical processes in extreme environments, including shock waves, jets, accretion discs, and X-ray coronae. Many of these vary on short time scales. However, current high-energy broad-band observations are still limited by the lack of high angular resolution, sensitive capabilities at energy ranges beyond Chandra's bandpass. Moreover, new multi-messenger facilities, both on the ground and in space, will increase the number of variable and transient objects by orders of magnitude in the upcoming decade. A high angular resolution broad-band X-ray mission will be critically needed by the astrophysics community to investigate the nature of these sources. The emphasis of this report is deliberately on the hard X-ray band. Missions with capabilities below ~10 keV are relatively abundant and will remain so through the next decade, whereas focusing capabilities above ~10 keV are scarce and have limited capabilities currently making >10 keV observations the primary gap for broad-band X-ray science.

This report lays out the current landscape of high-energy facilities (§2), including both current and planned missions, and then describes a few key science drivers for broad-band X-ray astrophysics (§3). Section 4 discusses the current technology landscape, including both optics and detectors. In short, broad-band X-ray astrophysics is ripe for scientific exploration and is able to address a range of critical questions highlighted by the Astro2020 Decadal Survey. The largest efforts to date with focusing optics have not reached beyond the Small Explorer class, while technological advances enable a low-risk architecture at the Probe or Flagship class that achieve order of magnitude or greater improvements in capabilities.

## 2. Current Landscape

We briefly review the current landscape of high-energy astrophysics missions with focusing optics working above ~12 keV. Non-focusing telescopes, such as the Insight-HXMT (Zhang et al. 2020) or coded aperture telescopes such as AstroSat/CZTI (Bhalerao et al. 2017) and Swift/BAT (Barthelmy et al. 2005), are not discussed as such technologies, though able to efficiently survey the sky, are inherently low angular resolution and have limited sensitivity. The plethora of soft X-ray facilities capable of focusing observations at <10 keV are also not discussed here since they are well documented elsewhere and are not the primary limiting factor for a broad-band X-ray observatory. We begin with currently operating missions (§2.1), followed by approved missions in development (§2.2), and proposed missions under review (§2.3). We do not discuss concepts that were not selected, such as the High-Energy X-ray Probe (HEX-P; Garcia et al. 2024, Madsen et al. 2024, Stern et al. 2025) and Simbol-X (Ferrando et al. 2006), facilities that are no longer in operation (e.g., Hitomi/HXI; Nakazawa et al. 2018), or concepts being developed for future opportunities. This report leans heavily on the significant efforts by the community for this full gamut of missions and mission concepts.

## 2.1 Operating Missions

**NuSTAR.** The Nuclear Spectroscopic Telescope Array (NuSTAR; Harrison et al. 2013) launched in 2012 and was the first orbital telescope capable of focusing high-energy "hard" X-rays (3 to 79 keV). NuSTAR provides images that are 10 times sharper and 100 times more sensitive than previous instruments. Designed as a two-year NASA Small Explorer (SMEX) mission with observations planned by the science team, NuSTAR has been operating for over a decade as a guest observatory with targets competitively selected through an annual call for proposals with typical (and rising) oversubscriptions of >3X overall and >6X for targets-of-opportunity (ToOs). The NuSTAR angular resolution, measured as the half-power diameter (HPD), is 1 arcminute, and though not designed for transients, recent updates to scheduling and ground communications enable NuSTAR to execute a rapid-response ToO observation in less than 6 hours using the new automated response system. Currently, >25% of NuSTAR observations are for ToOs, >50% of observations are coordinated with other (generally low-energy X-ray) facilities, and >70% of observations are time-constrained.

**SRG/ART-XC.** The only other focusing hard X-ray telescope in operations currently is the Russian-led Mikhail Pavlinsky Astronomical Roentgen Telescope – X-ray Concentrator (ART-XC; Pavlinsky et al. 2021) onboard the Spectrum-Roentgen-Gamma (SRG) satellite. Launched in 2019, ART-XC maps the sky at 4–30 keV. While the detector pixel size is ~0.75 arcmin, the mirror modules provide angular resolution as low as 25 to 30 arcsec (HPD) on-axis. When convolved with the size of the pixels in the focal plane, the angular resolution of ART-XC is approximately 0.5 arcmin (HPD) for typical survey mode observations that combine on- and off-axis data. Importantly, access to data from ART-XC is currently largely limited to the Russian high-energy community.

## 2.2 Missions in Development

**SuperHERO.** The Super-High Energy Replicated Optics (SuperHERO; Thomas et al. 2025) mission is a NASA-funded, high-altitude balloon-borne hard X-ray telescope observatory with its inaugural two-day science flight scheduled for Fall 2029 from Fort Sumner, New Mexico. SuperHERO is designed to achieve an angular resolution of <10 arcsec (HPD) in the 20-40 keV range.

## 2.3 Missions under Review

**EXPO.** The Enhanced X-ray Polarimetry Observatory (EXPO; Soffitta et al. 2026) is a mission concept proposed to ESA as an M8 medium-class candidate in early 2026, with a prospective launch around 2041. EXPO uses focusing optics to provide polarimetric information in the 2-35 keV energy range.

**PHEMTO.** The Polarimetric High Energy Modular Telescope Observatory (PHEMTO; Virgilli et al. 2026) is another mission concept proposed as an ESA M8 candidate. Using a combination of focusing and Laue optics, PHEMTO promises high angular resolution broad-band X-ray observations out to 600 keV.

The M8 mission selection process is currently in the proposal evaluation and scientific ranking phase, with the selection of up to five candidate missions for formal study expected to be announced in November 2026.

## 3. Science

Broad-band X-ray observations enable a broad swath of unique and important science not accessible from soft X-ray observations alone. For the sake of brevity, we detail just two science cases below: one extragalactic science case (§3.1), understanding the summed emission from cosmic X-ray sources, which is dominated by accretion onto supermassive black holes; and one Galactic science case (§3.2), understanding accretion onto stellar mass compact objects (e.g., black holes and neutron stars) in our Galaxy. However, broad-band X-ray observations provide key data to address a multitude of high-energy questions. Table 1 lists refereed papers formulated in support of the HEX-P mission proposal in 2023-2024, though, by design, these papers address important, unique science made possible with a generic, sensitive, broad-band X-ray observatory and thus relevant to this enterprise. We also provide Table 2, which lists the dates and topics for meetings of the BBX-SAG between April and June 2026.

**Table 1.** Publications describing broad-band X-ray science, prepared for the HEX-P mission concept.

| Publication | Topic |
|---|---|
| Alford et al. (2024) | Magnetars and isolated neutron stars |
| Bachetti et al. (2023) | Ultraluminous X-ray sources (ULXs) |
| Boorman et al. (2024) | Obscured active galactic nuclei |
| Brightman et al. (2024) | Transient phenomena in the 2030s |
| Civano et al. (2024) | Cosmic X-ray background |
| Connors et al. (2024) | Accretion onto stellar mass black holes |
| Kammoun et al. (2024) | Physics of the X-ray corona |
| Lehmer et al. (2023) | Nearby galaxy surveys |
| Ludlam et al. (2023) | Accretion onto neutron stars |
| Marcotulli et al. (2024) | Blazars and jets |
| Mori et al. (2023) | PeVatrons, superbubbles, microquasars, and gamma-ray binaries |
| Mori et al. (2024) | Galactic center: Sgr A*, compact object binaries, and diffuse emission |
| Pfeifle et al. (2024) | Dual active galactic nuclei |
| Piotrowska et al. (2024) | Supermassive black hole spin |
| Reynolds et al. (2023) | Supernova remnants, pulsar wind nebulae, and nuclear astrophysics |

**Table 2.** Presentations to the BBX-SAG.

| Date | Lead | Topic |
|---|---|---|
| 2026 Apr 9 | Chen, Madsen, Stern | Kick-off meeting |
| 2026 Apr 16 | Wilkins | Science: supermassive black holes |
| 2026 Apr 30 | Bongiorno, Madsen | Technology: mirrors |
| 2026 May 7 | Garcia | Science: Galactic binaries |
| 2026 May 14 | Miller, Hagino | Technology: detectors |
| 2026 May 21 | Wik | Science: galaxy clusters |
| 2026 May 28 | Gurgew | Technology: coatings |
| 2026 June 4 | Stern | Science: time domain |

## 3.1 The Cosmic X-ray Background (CXB)

The diffuse cosmic X-ray background (CXB; Figure 1) was discovered before any individual extragalactic X-ray source, and even before the cosmic microwave background. The 1962 Aerobee rocket flight by Giacconi and collaborators found not only the first extra-solar X-ray source, Sco X-1, but also an apparently isotropic X-ray glow that could not be attributed to any single object. Before focusing X-ray observatories, its nature remained one of the defining puzzles of high-energy astrophysics: was it truly diffuse, or the integrated light of a vast population of discrete, unresolved sources? The answer, now firmly established, is that the CXB is overwhelmingly the summed emission of accreting supermassive black holes — active galactic nuclei (AGN) — integrated over cosmic history. Its spectral shape and intensity encode the demographics of that population: how many black holes were growing, at what rate, behind how much obscuring gas, and at what epoch.

The breakthrough came from deep imaging. The Chandra Deep Fields and complementary XMM-Newton surveys resolved the soft-band (~0.5–2 keV) CXB into discrete sources at the ~80–90% level, and the 2–8 keV band comparably, exploiting XMM-Newton's large effective area and Chandra's sub-arcsecond resolution. The resolved sources are overwhelmingly AGN, spanning a wide range of luminosity, obscuration, and redshift. Joined with multiwavelength coverage from the optical, infrared, and radio, these surveys became the empirical backbone of modern galaxy evolution: they established the strikingly parallel cosmic history of black hole accretion and star-formation, revealed the black-hole–host coevolution implied by the local M–σ relation, and anchored the picture in which AGN feedback regulates star formation in massive galaxies. The questions that motivate the field today — how black holes are fueled, how accretion energy couples back to the host, and when the most obscured growth phases occur — all rest on this foundation of resolved source counts.

And yet the accounting is conspicuously incomplete where it matters most. The CXB peaks in intensity near 20–30 keV, yet it is precisely above 10 keV — where the bulk of the energy density

lives — that the resolved fraction collapses. Hard X-ray imaging missions such as NuSTAR have made the first direct surveys of this band, but their limited effective area and modest angular resolution have resolved only a small fraction of the >10 keV background, leaving the bulk of the CXB largely unaccounted for at the source level.

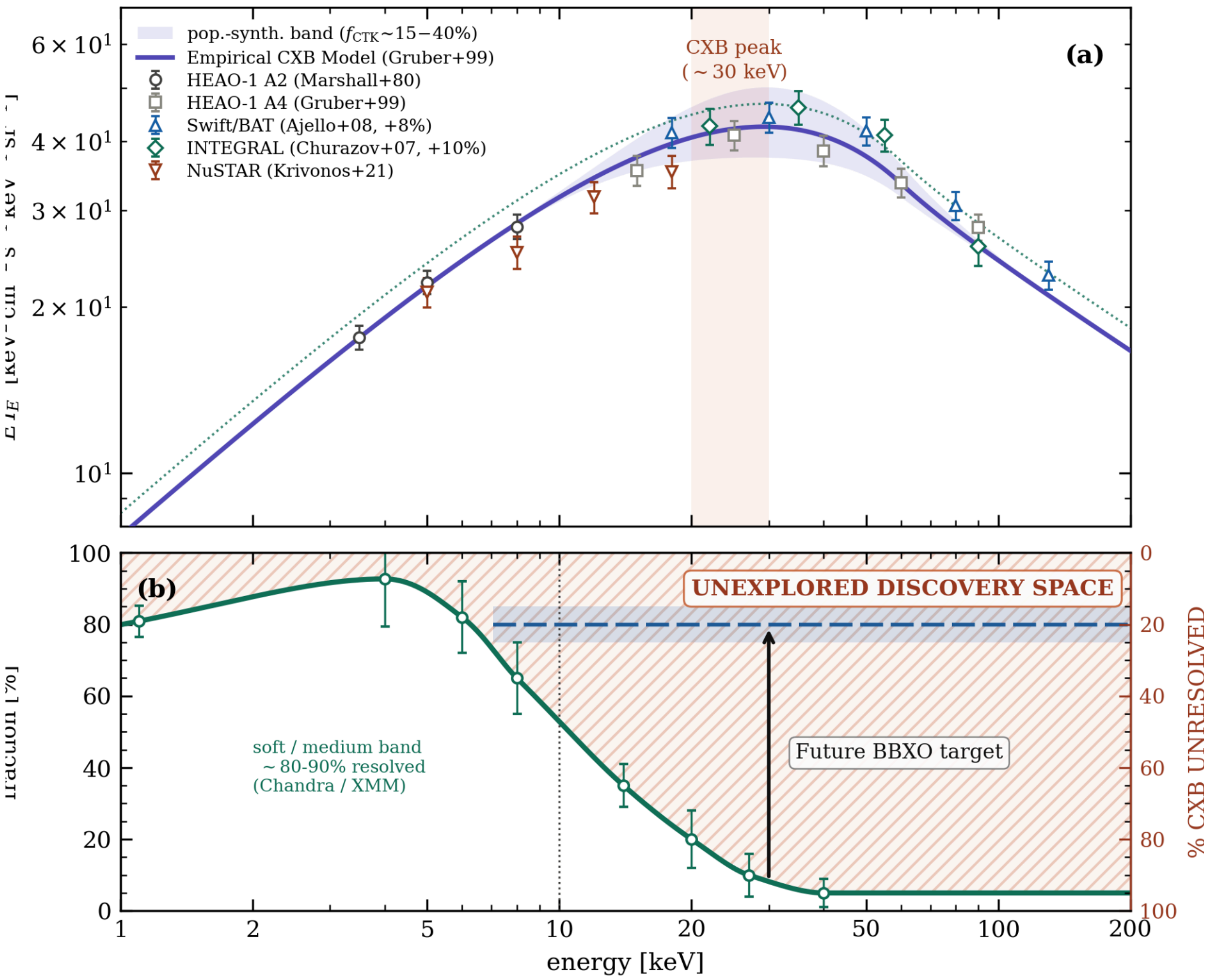

**Figure 1.** [*Top*] Spectrum of the cosmic X-ray background (CXB) measured by multiple missions over the past 40 years, showing the peak at ~30 keV. The purple band spans population-synthesis models with a wide range of Compton-thick AGN fractions, which rely on assumptions about AGN spectral and population models that a broad-band X-ray observatory (BBXO) would be needed to address. [*Bottom*] Fraction of the CXB that is resolved into individual sources as a function of energy. While Chandra and XMM-Newton have resolved most of the CXB at lower energies (<7 keV), the CXB is largely unresolved at higher energies. At its ~30 keV peak, only ~10% of the CXB is resolved into individual sources.

This gap hides some of the most fundamental questions in galaxy evolution. Population synthesis models (e.g., Ananna et al. 2019) reproduce the CXB by summing AGN over an assumed distribution of luminosities, spectral shapes, absorbing columns, and redshifts. Matching the 20–30 keV CXB peak requires a substantial contribution from the reflection spectral component from the heavily obscured and Compton-thick AGN with obscuration column densities exceeding ~$10^{24}$ cm$^{-2}$, strongly attenuated below 10 keV and transmitted only in the hard band. Locally, confirmed Compton-thick detections remain few; at cosmic noon and earlier, existing CXB measurements cannot distinguish models spanning a wide range of Compton-thick fraction. Because soft surveys are nearly blind to these sources, the discriminating data simply does not exist at the required depth. This matters because the obscured population sets the total accretion-driven energy budget across cosmic history, while the obscured-to-unobscured ratio constrains the duty cycle and geometry of black hole fueling — both of which determine how much energy is available for feedback.

Recent discoveries sharpen the stakes. JWST has uncovered an abundant population of compact, red sources at $z \gtrsim 4$ — the "little red dots" — many showing spectroscopic signatures of accreting black holes yet remarkably X-ray weak in existing soft-band data. Whether they are intrinsically X-ray faint, extremely obscured, or accreting in an exotic regime remains unresolved. This highlights the necessity of a broad-band X-ray understanding of obscured black-hole accretion — from the heavily absorbed AGN of the local universe, where hard X-rays penetrate the obscuring material, to the earliest quasars, where the same bandpass reaches the rest-frame coronal cutoff and the physics of the accretion flow itself.

The implication for instrumentation is direct. Resolving and understanding the CXB requires extending the X-ray observation capability into a broad spectral coverage spanning soft and hard bands in a single instrument, sharp angular resolution to overcome source confusion, and the sensitivity to reach the populations that dominate the unresolved peak. A broad-band imaging X-ray observatory is therefore not an incremental improvement but the specific capability needed to complete the accounting of cosmic black hole growth — to convert the integrated glow of the CXB back into the resolved population that produced it, across the full range of obscuration and redshift, and in particular for the obscured and high-redshift sources that soft-band surveys, by construction, cannot see.

## 3.2 Galactic Science Cases: X-ray Binaries and More

Compact Galactic sources and their counterparts in nearby galaxies — black-hole and neutron-star X-ray binaries, accreting white dwarfs, ultraluminous X-ray sources (ULXs), supernova remnants and pulsar-wind nebulae, and a broad range of transient phenomena — are among the most informative laboratories available for studying accretion physics, the dense-matter equation of state, strong-field gravity, magnetic fields near the quantum-electrodynamics limit, stellar-mass compact-object demographics, and particle acceleration. Their defining observational property is that the physics of interest is encoded across the full X-ray band, from below 1 keV to roughly 100 keV, and that the relevant systems vary on timescales from milliseconds to months.

Considering X-ray binaries, i.e., a compact object such as a neutron star or black hole accreting from a stellar companion, X-rays are radiated across the entire ~0.1–100 keV band, motivating a broad-band facility. Thermal emission from the accretion disk peaks below ~2 keV. Relativistic reflection imprints the Fe K emission complex at 6–7 keV, the Fe K absorption edge at 7–8 keV, a Compton hump at ~20–30 keV, and the coronal continuum that turns over at higher energies. State transitions, line-of-sight obscuration, and coronal high-energy cutoffs all manifest above the natural ~10 keV “limit” where current soft-band missions lose sensitivity. For neutron star X-ray binaries, cyclotron resonant scattering features fall between 5 and 100 keV and provide the only direct probe of those field strengths. Furthermore, because these sources evolve on timescales from milliseconds to months, the broadband components must be measured at the same time to maximize their value for physical understanding of the system. Simultaneous broad bandwidth observations are essential.

Below we highlight a few Galactic and nearby-galaxy topics, tying the science to concrete capability drivers. These questions are intended to populate a traceability matrix in which measurement drivers can be ranked and the cases for which hard-band capability is mission-critical (rather than merely beneficial) can be identified.

**Stellar-mass black hole accretion.** With roughly thirty confirmed black hole binaries in the Galaxy, evolving through outbursts on timescales of weeks to months, these systems are the cleanest laboratory for studying accretion physics and the coupling between the accretion disk, the Comptonizing corona, and the relativistic jet. X-ray reflection spectroscopy yields the black hole spin, the inner-disk radius, the height of the corona, and the iron abundance. Simulations for the HEX-P concept show that reflection can be constrained even in the faint hard state, at luminosities below 0.1% of the Eddington luminosity — precisely the regime in which present-day missions lose the Compton hump (Connors et al. 2024). Joining the soft and hard telescopes in a single simultaneous fit breaks the degeneracies between continuum and reflection that cannot be resolved with a hard-band instrument alone (see the case of EXO 1846-031 presented in Connors et al. 2024). The corresponding capability driver is simultaneous coverage from roughly 0.2 to 80 keV at high count rate, with pile-up control and a high-rate telemetry mode for bright Galactic outbursts.

**Neutron star equation of state.** Reflection sets a hard upper limit on the neutron-star radius: the inner-disk radius obtained from a broadband fit provides a physical bound on the stellar radius. A key limitation of current NICER-plus-NuSTAR fits is that they require the spin of the neutron star to be fixed at zero, which biases the inferred radii. Sufficient broadband sensitivity enables reflection fits with the spin left free, tightening the mass–radius constraint and complementing both NICER pulse-profile modeling and gravitational-wave tidal measurements (Ludlam et al. 2023). The result is a direct discriminator between hadronic, hyperon, and quark-matter equations of state. The capability driver is sensitivity to free neutron-star spin in joint reflection–continuum fits, which in turn demands a stable continuum from the soft thermal component, through Fe K, into the 20–80 keV band.

**Neutron star magnetism.** Magnetic field strengths in accreting neutron stars span from about $10^{12}$ G in high-mass X-ray binaries (HMXBs) to roughly $10^{15}$ G in magnetars. Cyclotron resonant scattering features between 5 and 100 keV are the only direct probes of these fields, with the line energy related to the field strength by $E_{cyc} \approx 11.6 \times (B / 10^{12}\ G)$ keV. Magnetic neutron stars also drive super-Eddington accretion in ULXs: the pulsating ULXs M82 X-2, NGC 5907 ULX-1, and NGC 7793 P13 reach luminosities of order 100 times Eddington, and the cyclotron line in M51 ULX-8 (Brightman et al. 2018) suggests a field of $10^{14}$–$10^{15}$ G. Magnetars and isolated neutron stars probe the vacuum-QED limit and show hard-tail evolution after outbursts. The capability driver is simultaneous 5–80 keV coverage with timing modes for phase- and luminosity-resolved cyclotron physics, across the full field range from HMXBs through magnetars to ULXs.

**White dwarfs and the Galactic Ridge.** Accreting white dwarfs — in particular, magnetic cataclysmic variables and their post-shock plasmas — and the origin of the unresolved hard emission of the Galactic Ridge are addressed through the shape of the bremsstrahlung cutoff and the 6.4–6.97 keV iron complex. The capability driver is 10–50 keV imaging at high angular resolution, which is what turns the integrated ridge emission into a population of resolved sources and constrains the relative contributions of source classes. Importantly, hard X-ray data provides masses for any of those sources that do turn out to be magnetic white dwarfs, eliminating the need for optical spectroscopy.

**Compact object populations.** Below 10 keV the neutron-star and black-hole populations overlap strongly, and the hard, soft, and intermediate spectral states are degenerate in soft-only data. Adding the 12–25 keV band cleanly separates compact-object types, assuming hard-band imaging that beats the source confusion that limits NuSTAR in crowded fields. Together, these capabilities will enable the first sensitive hard-X-ray census across a range of metallicity, star-formation rate, and stellar age (Lehmer et al. 2023), giving direct constraints on the formation channels of high-mass and low-mass X-ray binaries. The capability driver is hard-band imaging above 10 keV for nearby galaxies; for this science it is a requirement, not a luxury.

**Galactic Center and nearby galaxy demographics.** NuSTAR detected 66 sources above 10 keV near the Galactic Center but was confusion-limited; HEX-P simulations indicate that more than a thousand sources could be detected in the 2–20 keV band within 30′ of Sgr A* with a Probe-class mission, and that the diffuse hard emission could be resolved into either a bound population of black-hole binaries or a population of central pulsars (Mori et al. 2024). For nearby galaxies — M31, Maffei 1, NGC 253, NGC 3310 — surveys would detect more than two hundred X-ray binaries each, with over 130 detected above 12 keV, spanning roughly an order of magnitude in metallicity and a diverse range of star-forming environments (Lehmer et al. 2023). The capability driver is a sub-10″ half-power diameter above 10 keV, which sets the confusion floor both for the densest stellar field in the Galactic Center and for resolved studies in galaxies at a few megaparsecs.

**Galactic particle accelerators.** Hard X-rays are a direct probe of particle acceleration. The synchrotron cutoff in the 10–80 keV band constrains the maximum particle energy, the limits of acceleration, and transport; pairing hard X-rays with TeV gamma rays and neutrinos disentangles hadronic from leptonic scenarios. The $^{44}$Ti line at 78 keV in young supernova remnants constrains explosion asymmetry and nucleosynthesis through its morphology and line shape, and resolving the substructure of pulsar-wind nebulae and remnants at hard energies adds information beyond the integrated flux (Mori et al. 2023; Reynolds et al. 2023; Grefenstette et al. 2014, 2017). The capability driver is 10–80 keV imaging at low background, with mosaic survey capability for extended-source morphology and narrow-line sensitivity near 100 keV.

**Transient X-ray binaries.** Outburst evolution sets the required science cadence: a source can move from the hard rise, through intermediate states, to the soft state and back to a hard state within weeks, and each state probes a different reflection geometry, corona, and degree of disk truncation. Engine-powered and embedded transients — fast blue optical transients such as AT2018cow, embedded supernovae, gravitational-wave counterparts, and magnetar outbursts — are often first detectable in hard X-rays, which penetrate circumstellar material that soft X-rays cannot, leaving only an ambiguous continuum and weak hints of Fe K emission at lower energies. A HEX-P simulation of AT2018cow, based on the NuSTAR observations of Margutti et al. (2019), reveals reflection from a newly formed central engine (García et al. 2024). Capturing a full broadband spectrum at every visit, with no need for co-orbital coordination, requires rapid (< 24 h) ToO turnaround, high observing efficiency, and a large field of regard; here hard-band sensitivity and operational agility are coupled.

**Key capability trade space.** The science cases above motivate a small set of candidate requirement axes:

> *Energy range:* a 0.2–80 keV baseline, testing the specific value of coverage at or above 80 keV for high-energy cutoffs, cyclotron lines, and nuclear lines.
>
> *Hard-band angular resolution*: the parameter that sets the confusion floor in the Galactic Center, in nearby galaxies, and in star-forming regions.
>
> *Throughput and background*: the drivers of hard-band source classification, faint non-thermal morphology, and ToO spectroscopy.
>
> *Timing and count-rate handling*: required for pulsations, quasi-periodic oscillations, bursts, magnetar tails, and bright Galactic outbursts.

*Field of view and survey strategy*: needed for mosaics of the Galactic Center, Galactic Ridge and nearby galaxy studies.

*Agility and operations*: rapid ToOs, observing efficiency, target visibility, and a community-trigger model.

Each one of these axes can be evaluated against four criteria: (1) is the capability *unique*; (2) how many science cases it enables; (3) how sensitive the science is to its degradation; and (4) its operational complexity. The final goal is to produce a ranking of what capabilities must absolutely be preserved and identify where trades are acceptable.

**Summary.** Compact Galactic sources span a wide range of extreme regimes — strong-field gravity around stellar-mass black holes, the dense-matter equation of state in neutron stars, magnetic fields approaching the quantum-electrodynamics limit, particle acceleration in supernova remnants and pulsar-wind nebulae, and a growing population of multi-messenger transients — and across all of them the decisive measurements are distributed across the soft and hard X-ray bands and must be made simultaneously.

The science cases assembled here converge on a single conclusion: broad energy coverage is necessary but not sufficient. Realizing this science also requires hard-band angular resolution that beats source confusion in crowded fields, throughput and background control for faint non-thermal emission and surveys, timing and count-rate performance that preserves the broadband fit on bright sources, and the operational agility to respond to Galactic and extragalactic transients. These are best treated as mission-enabling requirements rather than optional enhancements. The traceability exercise proposed here— scoring each capability by its uniqueness to BBX, the number of science cases it enables, the sensitivity of the science to its degradation, and its operational complexity — is the mechanism for converting this list into ranked, quantitative requirements, and for distinguishing the capabilities that a mission concept must protect from those where trades are acceptable.

# 4. Technology

The Broadband X-ray Observatory (BBXO) architecture requires careful consideration of multiple competing design parameters. The requirement for low grazing incidence angles, necessitated by the physics of hard X-ray reflection, drives the need for long focal lengths. But small mirror radii, while beneficial for achieving good high energy response, limit the collecting area of individual mirrors and lead to the use of multiple modules as the only way to build up effective area. Additionally, while longer focal lengths increase the efficiency of the optic and increase the bandpass, it reduces the accessible field of view, limiting survey efficiency. The optimized BBXO architecture is therefore a balance between science requirements, physics, and technology.

## 4.1 Detectors

For a balanced, general purpose observatory, a recommended Field of View (FoV) of 20’ at a 20 meter focal length will require a detector plane of ~120 mm, and for 30’ at a focal length of 10 m a detector plane of ~90 mm. Typical linear sizes of hard X-ray detectors are about 20-40 mm, which requires most detector planes to be tiled. For reasonable FoVs, focal lengths longer than 20 meters will result in very large and challenging focal planes, which coupled with multiple mirror modules, would result in an unfeasibly large number of detectors. The cap on the focal length is therefore primarily limited by the detector technology.

Table 3 shows a representative sample of hard X-ray detector technologies currently in use and under study. There is no obvious choice of detector for BBXO at this time. As described below, each detector excels in different areas with different challenges, and largely the choice is between CdTe and CZT detectors of various type and geometry, distinguishing themselves in cost and availability – commercial CdTe types with worse energy resolution versus bespoke CZT detectors with low noise electronics and high crystal quality – and actively cooled high energy resolution detectors. Below we briefly outline each type.

**Table 3.** Representative detector technologies.

| Parameter | CdTe DSD | HEXITEC | Germanium | Gamma-ray TES | NuSTAR CZT | Redlen CZT |
|---|---|---|---|---|---|---|
| Material | CdTe | CdTe (1 mm thick) | Ge | Sn absorber | CZT (2mm thick) | CZT (3mm thick) |
| Energy Resolution | 0.7-2 keV | 0.8-1.2 keV | 120 eV @ 6 keV; 300 eV @ 60 keV | 55 eV @ 97 keV | 0.95 keV @ 60 keV | 0.55 keV @ 60 keV |
| Pixel Pitch | 60-100 μm | 250 μm | ≥8 μm | 1.3 mm | 0.6 mm | 0.6 mm |
| Energy Range | 1-150 keV | 1-150 keV | 1-100 keV | 1-200 keV | 2-150 keV | 2-200 keV |
| TRL | 9 | 4-5 | 4 | 4 | 9 | 4 |
| Flight Heritage | Hitomi FOXSI | No | No | No | NuSTAR | No |
| Key Advantages | Mature, customi-zable, compact | High readout rate, moderate resolution | Very high energy resolution | Exceptional energy resolution | Flight proven, high energy resolution | 40% better energy resolution than NuSTAR, no cooling required |
| Key Challenges | Lower energy resolution | Not yet flown | High dark current, cooling required | Slow, cryogenics, pixel gaps, thermal uniformity | Low yield | Not yet flown |
| Development Status | Operational | Testing complete | Early development | Proposed for ASCENT | Operational | Under detailed characteri-zation |

**CdTe DSD.** CdTe double-sided strip detectors (DSDs) are a mature technology with a TRL of 9 and are commercially available. These detectors are constructed from cadmium telluride (CdTe) material and offer moderate energy resolutions of approximately 1 keV for silicon and 2 keV for CdTe on the HXI instrument, with FOXSI achieving 0.7-0.8 keV. The detectors feature a compact 32 × 32 mm footprint with pixel pitches ranging from 60 to 100 microns. They have been successfully deployed on Hitomi and FOXSI, and are also being utilized for the SuperHERO mission with increased detector thickness. The technology employs VATA ASIC readout electronics and can be configured with silicon DSD layers on top of the CdTe, offering flexibility for customization while maintaining reduced costs due to commercialization efforts.

**HEXITEC.** HEXITEC detectors utilize 1 mm thick CdTe material and demonstrate excellent energy resolution with 0.8 keV at 60 keV and 1.2 keV at 159 keV. These detectors have a 40 × 40 mm active area divided into 80 × 80 pixels with a 250-micron pitch. Currently at TRL 4-5, HEXITEC detectors are being developed through work at NASA Goddard Space Flight Center but have not yet been flown in space. They were proposed for a solar SMEX and underwent rigorous testing at a heavy ion facility to evaluate their radiation hardness. The primary design driver for HEXITEC is achieving high readout rates, making them particularly suitable for high-flux applications.

**Germanium-based detectors.** Germanium offers energy resolution comparable to silicon CCDs, representing one of the highest-resolution options available for hard X-ray detection. These detectors are available in 200 mm wafer formats with pixel pitches as small as 8 microns. Currently at TRL 4, germanium detectors are being developed at MIT Lincoln Laboratory and were featured in the ROSES-24 APRA. Development efforts are focused on fabricating a variety of pixel arrays, as large as 512 × 512 pixel framestore devices with a 24-micron pitch. While offering exceptional energy resolution of approximately 300 eV, germanium detectors present significant challenges including high dark current that necessitates active cooling to ≤-150°C. Their operational efficiency is limited to energies up to 100 keV, where they achieve ~20% detection efficiency at 80 keV.

**Gamma-ray TES.** High Resolution Hard X-ray Transition Edge Sensor (TES) microcalorimeter arrays utilize tin (Sn) absorbers to achieve extraordinary energy resolution of 55 eV at 97 keV across a 225-pixel array. Currently at TRL 4, these detectors benefit from established fabrication capabilities at both NIST and NASA Goddard Space Flight Center. The technology has been proposed for the ASCENT mission and operates over a broad energy range from 1 to 200 keV. Current implementations feature 1.3 mm pixels, with typical arrays containing approximately 200 pixels, though scaling to larger formats presents significant challenges due to thermal uniformity requirements. The primary limitations include slow pulse processing with 10 millisecond time resolution and the requirement for cryogenic operation. Additionally, the detectors face geometric constraints as individual pixels cannot be placed in direct contact, necessitating gaps between pixels that reduce the overall fill factor and detection efficiency.

**CZT.** The NuSTAR CZT hybrid detectors have been operating successfully on NuSTAR since 2012. Developed at Caltech's Space Radiation Lab over the past two decades, these detectors combine carefully screened 2 mm thick CZT crystals produced by eV Products with a custom low-noise, low-power ASIC designed at Caltech. Each detector consists of a 32 x 32 array of 0.6 mm pixels covering 20 x 20 mm. These detectors achieve a spectral resolution of 0.95 keV at 60 keV. They have background rejection coming from depth-of-interaction measurements which also enable spectral corrections at high energies. There are two primary limitations of these devices — the quantum efficiency drops to 40% at 150 keV, and the ASIC caps the readout at 400 cps/module. To overcome some of these limitations, Caltech is currently developing and characterizing a new generation of CZT hybrids using 3 mm thick crystals produced by Redlen Technologies (50% thicker than NuSTAR flight detectors). While these devices retain the proven 32 x 32, 0.6 mm pixel format with the existing ASIC readout, the increased thickness substantially improves stopping power, quantum efficiency, and spectral resolution in the 2-200 keV band. Ongoing characterization efforts at Caltech SRL have shown a spectral resolution of 0.55 keV @ 60 keV, roughly 40% better than NuSTAR-era devices.

## 4.2 Optics

Large aperture optics, such as the mirror baselined for NewAthena (one structure 2.4 m across), are not well suited for hard X-ray response. Hard X-rays require grazing incidence angles of mrad, not achievable with large radius shells, and the design therefore favors densely packed small radius optics. However, if the radius becomes too small and the angle too shallow, straylight from single bounce photons becomes an issue, contaminating the FoV. An optic of fixed inner and outer radius will always perform better at longer focal lengths, and so from the point of view of the optic, as long a focal length as possible is desired. The inner radius is set by the geometry to suppress stray light, and the outer radius as a trade between low energy effective areas; this is a mass trade between number of modules vs individual module size.

There are five viable options for hard X-ray mirror technology. Each is optimized for either throughput (dense packing of very thin shells), weight (large aperture optics), or angular resolution. Below we discuss each in turn:

**Replicated optics.** Replicated optics technology delivers moderate effective areas through the use of multiple optical modules, with angular resolutions ranging from moderate to high quality (5"-20" depending on the specific design). This mature technology has been successfully deployed on numerous missions including FOXSI, ART-XC, IXPE, BeppoSAX, XMM, eROSITA, and eXTP. Two primary manufacturers have established expertise in this field: NASA's Marshall Space Flight Center (MSFC) specializes in smaller to moderate radius optics achieving 5"-20" resolution, while Media Lario in Italy focuses on larger radius optics with 8"-15" angular resolution. The technology offers a proven balance between performance and manufacturability and lends itself extremely well to multiple optics, since multiple shells can be replicated from the same mandrel. The main drawback is mass; although multiple optics can be produced easily, the resulting mirror assemblies are relatively heavy.

**Silicon pore optics.** Silicon pore optics (SPO) represents an advanced technology funded by the European Space Agency (ESA) specifically for the NewAthena Flagship mission. This technology achieves very large effective areas exceeding 1 square meter, though it delivers moderate angular resolution of approximately 9-10 arcseconds. The innovative design utilizes silicon's excellent material properties to create lightweight, high-throughput X-ray optics. The restriction on these optics is the inner radius. The modules are produced by bending Si plates and this approach is not well suited for small radius modules.

**Segmented mono-crystalline silicon optics.** Segmented mono-crystalline silicon optics represent cutting-edge technology offering high angular resolution better than 5 arcseconds. This technology leverages the exceptional flatness and stability of mono-crystalline silicon, segmenting them into precisely aligned mirror modules. There are no current limitations to this technology, but there are several challenges that would need to be overcome for hard X-rays to achieve dense packing, long shells, and small radii.

**Nested foil mirror optics.** Nested foil mirrors provide moderate to high effective areas while producing larger point spread functions of approximately 1 to 1.5 arcminutes. This approach is characterized by high throughput and cost-effectiveness, making it an attractive option for budget constraints or prioritization of photon collection over imaging sharpness. The technology has demonstrated its reliability through successful deployment on numerous missions including BBXRT, ASCA, Astro-E/Suzaku, InFOCuS, Hitomi (ASTRO-H), NICER, and XRISM. The thin foil construction allows for tightly nested configurations that maximize collecting area within constrained volumes, though the manufacturing process trades some optical precision for simplicity and reduced cost.

**Slumped glass.** Slumped glass, as used on NuSTAR, offers moderate to high effective areas with angular resolutions ranging from moderate to large (0.5 to 1 arcminute). Similar to nested foils, slumped glass optics emphasize high throughput and cost-effectiveness, making them suitable where collecting area is prioritized over angular resolution. The technology has been successfully implemented on the NuSTAR mission. The slumping process involves heating flat glass sheets and allowing them to conform to precisely shaped mandrels, creating lightweight mirror segments that can be assembled into nested configurations. This manufacturing approach provides an excellent balance between performance, weight, and cost for moderate-resolution X-ray imaging applications.

| | Effective Area | Angular Resolution (PSF) | Key Characteristics | Missions/ Applications | Manufacturer/ Developer |
|---|---|---|---|---|---|
| **Replicated Optics** | Moderate (multiple optics) | Moderate to high (5"-20") | Moderate effective areas with good angular resolution | FOXSI, ART-XC, IXPE, BeppoSAX, XMM, eROSITA, eXTP | MSFC (5"-20" smaller/moderate radius); Media Lario, Italy (8"-15" larger radius) |
| **Silicon Pore Optics** | Very large (>1 m²) | Moderate (~8") | ESA-funded for Athena flagship; very large collecting area | NewAthena | ESA development with Cosine (NL) |
| **Segmented Mono-crystalline Silicon** | Large | High (<5") | Large effective area with excellent angular resolution | Advanced X-ray missions | Research/development GSFC |
| **Nested Mirror Foils** | Moderate to high | Large (1'-1.5') | High-throughput, inexpensive X-ray optics | BBXRT, ASCA, Astro-E/Suzaku, InFOCuS, Hitomi (ASTRO-H), NICER, XRISM | Standard manufacturing GSFC |
| **Slumped Glass** | Moderate to high | Moderate to large (0.5'-1') | High-throughput, inexpensive X-ray optics | NuSTAR, IAXO | Standard manufacturing, Columbia, INAF Brera, others |

## 4.3 Coatings

The focal length and mirror geometry are all required to enable hard X-ray reflection, but ultimately it is the coating which determines the bandpass.

**Pt and W coatings.** Tungsten/silicon (W/Si) and platinum/carbon (Pt/C) multilayers are the standard coatings for hard X-ray missions, with W/Si providing well-understood, reliable performance up to 70 keV until limited by the tungsten K-edge at 69.5 keV, while Pt/C extends coverage to 78 keV. Both are used on NuSTAR and the only technology development is not to the coatings themselves, but to implement stress-relief measures to mitigate stress induced deformation of the mirror substrates for the high angular resolution optical options.

**Ni coatings.** Nickel-based multilayer coatings avoid the high-energy absorption edges that limit W/Si and Pt/C, maintaining good reflectivity beyond 70-80 keV, though the nickel K-edge at 8 keV reduces low-energy performance. Ni is challenging to deposit due to its ferromagnetic properties, and due to its relatively low density, compared to Pt and W, very thick multilayer coatings are required, which compounds the stress issue. The vanadium addition in NiV/C coatings optimizes optical constants for improved performance compared to pure nickel. Ni coatings are still in early development and have yet to be demonstrated in a flight-like prescription.

**Ni hybrid coatings.** Hybrid coatings (Ni + Pt or W) combine nickel-based layers with platinum or tungsten layers to create optimized multilayer structures that overcome the limitations of single-material approaches. By strategically layering different materials, these coatings should be able to maintain good performance across the critical 8 keV nickel edge while preserving high-energy capabilities up to 70-80 keV. While more complex to fabricate than simple two-material multilayers, hybrid coatings offer the most promising solution to combine the known performance of W and Pt coatings with an energy response extending past 80 keV. However, like for Ni coatings, hybrid coating development is still in the early stages.

### 4.4 Straw-man instrument design

The choice of optics will be heavily weighted by cost and mass. Due to the need for multiple mirrors the options lean towards the replicated mirror technology and the slumped glass NuSTAR technology. For very high angular resolution at intermediate energies, mono-crystalline silicon still shows strong promise. The replicated technology gives you better angular resolution; however, the slumped glass technology yields a superior effective area at lower mass and cost. A typical “mirror module” will roughly add about 400 $cm^2$ at 6 keV, 300 $cm^2$ at 25 keV and 90 $cm^2$ at 70 keV for a focal length of 20 meters, pending the exact design and coating parameters. For high energy response >80 keV, hybrid Ni coatings show the most promise. This suggests that an assembly of three or more mirror modules is needed to address the science.

## 5. Summary

This report provides a resource motivating the broad range of science accessible from a high-energy observatory that extends from the traditional <10 keV “soft” range of facilities such as the flagship-class Chandra, XMM-Newton, and NewAthena missions, into the “hard” X-ray range of 10-100 keV. Table 4 presents an incomplete list of how broadband X-ray observations address a sampling of Astro2020 Key Questions. Focused observations of those higher energies have, to date, only been obtained from missions with arcminute-scale angular resolution such as balloon experiments and NuSTAR. We detail two science cases, one extragalactic and one Galactic, and provide references that detail a range of broadband X-ray science (Table 1). In short, broadband high-energy observations address a range of key science. We summarize the current state of the required technology, noting which are well established and which would benefit from additional development. Finally, we present a notional strawman mission design.

**Table 4.** How BBX addresses Astro2020 Key Questions

| Astro2020 Key Question* | How addressed by BBX |
|---|---|
| **What are the mass and spin distribution of neutron stars and stellar mass black holes?** [PoCOEP: Q1] | Study isolated neutron stars [Alford et al. 2024]<br>Study accretion onto neutron stars [Ludlam et al. 2023]<br>Study accretion onto stellar mass black holes [Connors et al. 2024] |
| **What powers the diversity of explosive phenomena across the** | Study ultra-high energy gamma ray sources [Mori et al. 2023]<br>Unique studies of diverse range of TDAMM [Brightman et al. 2024] |
| **electromagnetic spectrum?**<br>[PoCOEP: Q2] | Study magnetar and neutron star outbursts [Alford et al. 2024]<br>Supernova remnants and pulsar wind nebulae [Reynolds et al. 2023] |
| **How do some compact objects eject material in nearly light-speed jets, and what is that material made of?**<br>[PoCOEP: Q3] | Study blazars and jets [Marcotulli et al. 2024]<br>Determine source and acceleration mechanism of ultra-high energy gamma ray sources (e.g., PeVatrons) [Mori et al. 2023]<br>Determine contribution of ultra-high energy gamma ray sources to the cosmic ray and neutrino backgrounds [Mori et al. 2023] |
| **What seeds supermassive black holes and how do they grow?**<br>[PoCOEP: Q4] | Resolve the cosmic X-ray background [Civano et al. 2024]<br>Supermassive black hole spin survey [Piotrowska et al. 2024]<br>Understanding the physics of the corona [Kammoun et al. 2024]<br>Complete the census of obscured AGN [Boorman et al. 2024]<br>Find dual AGN in obscured systems [Pfeifle et al. 2024] |
| **What are the properties of dark matter and the dark sector?**<br>[PoC: Q2] | Search for exotic particle decay (e.g., axions) [Candón et al. 2025, Ruz et al. 2025] |
| **How will measurements of gravitational waves reshape our cosmological view?**<br>[PoC: Q4] | Identify counterparts to stellar-mass black hole mergers in obscured environments (e.g., AGN accretion disks) [Bartos et al. 2017, McKernan et al. 2019]<br>Identify counterparts to supermassive black hole mergers in obscured environments (e.g., galaxy mergers) [Milosavljević & Phinney 2005; Piro et al. 2022] |
| **How do supermassive black holes form and how is their growth coupled to the evolution of their host galaxies?**<br>[PoG: Q3] | Resolve the cosmic X-ray background [Civano et al. 2024]<br>Supermassive black hole spin survey [Piotrowska et al. 2024]<br>Understanding the physics of the corona [Kammoun et al. 2024]<br>Find IMBHs in ULX/HLX population [Bachetti et al. 2023]<br>Complete the census of obscured AGN [Boorman et al. 2024] |
| **How do habitable environments arise and evolve within the context of their planetary systems?**<br>[PoEASS: Q3] | Study X-ray stellar flares and their contribution to atmospheric escape [Hu et al. 2025] |
| **What are the most extreme stars and stellar populations?**<br>[PoSSSP: Q1] | Determine the nature of compact object binaries in the Galaxy and nearby galaxies [Lehmer et al. 2023, Mori et al. 2024] |
| **How does multiplicity affect the way a star lives and dies?**<br>[PoSSSP: Q2] | Determine the nature of compact object binaries in the Galaxy and nearby galaxies [Lehmer et al. 2023, Mori et al. 2024] |

***Astro2020 Panels:** PoCOEP = Panel on Compact Objects & Energetic Phenomena, PoC = Panel on Cosmology, PoG = Panel on Galaxies, PoEASS = Panel on Exoplanets, Astrobiology and the Solar System, PoSSSP = Panel on Stars, the Sun, & Stellar Populations